\documentclass[prd,twocolumn,groupedaddress,noshowpacs,nofootinbib]{revtex4}
\usepackage{graphicx}
\usepackage{dcolumn}
\usepackage{amssymb}
\usepackage{mathrsfs}
\usepackage{amsmath}
\usepackage{epsfig}
\usepackage[dvips]{color}
\usepackage{hhline}

\begin{document}

\title{Ultralight Dirac neutrinos from nearly degenerate Higgs doublets }

\author{Pei-Hong Gu}

\email{phgu@seu.edu.cn}

\affiliation{School of Physics, Jiulonghu Campus, Southeast University, Nanjing 211189, China}

\begin{abstract}

Two Higgs doublets respect a mirror symmetry with spontaneous violation so that their vacuum expectation values can realize a small difference. Under this symmetry, three newly introduced right-handed neutrinos rather than the standard model fermions perform an odd transformation. Accordingly the neutrino masses and the charged fermion masses respectively are proportional to the difference and sum of the vacuum expectation values of two Higgs doublets. From a phenomenological perspective, such nearly degenerate Higgs doublets with large cancellation are equivalent to a Dirac seesaw mechanism with high suppression.

\end{abstract}

%\pacs{98.80.Cq, 14.60.Pq, 95.35.+d, 12.60.Cn, 12.60.Fr}

\maketitle

\section{Introduction} 

The neutrino oscillation experiments have established that three flavors of neutrinos should be massive and mixing \cite{navas2024}. Moreover, the cosmological observations have indicated that the neutrinos should be extremely light \cite{navas2024}. On the other hand, the neutrinos are massless in the standard model (SM) \cite{navas2024}. In order to generate the neutrino masses, the most intuitive idea is to introduce three generations of right-handed neutrinos and then construct the Yukawa couplings of these right-handed neutrinos to the SM lepton and Higgs doublets. Then the neutrinos obtain their masses in a same way with the SM charged fermions. Unfortunately, such Yukawa couplings for the neutrino mass generation are too small to be naturally understood. Alternatively, people proposed the famous seesaw mechanism \cite{minkowski1977,yanagida1979,grs1979,ms1980,mw1980,sv1980,cl1980,lsw1981,ms1981,flhj1989,tao1996,ma2006} to elegantly guarantee the massive neutrinos ultralight. The essence of the seesaw mechanism is to efficiently suppress the neutrino masses by a ratio of a small physical quantity over a large physical quantity. 

In the most popular seesaw scenarios, the neutrino masses originate from certain lepton number violating interactions and hence the neutrinos have a Majorana nature. However, we should keep in mind that the theoretical assumption of the lepton number violation and then the Majorana neutrinos has not been confirmed by any experiments so far \cite{navas2024}. Therefore, it is worth studying the possibility of Dirac neutrinos. In analogy to the conventional seesaw models for generating the Majorana neutrino masses, we can construct some Dirac seesaw models \cite{rw1983,rs1984,mp2002,gh2006,gs2007,gu2012,cg2023} for generating the Dirac neutrino masses.

In the present work, we shall explore an unconventional approach to understand the neutrino mass generation. Firstly we shall consider a proper symmetry breaking to naturally realize a large cancellation between two physical quantities of the same magnitude. Then we shall resort to such cancellation for generating the ultralight neutrino masses. Specifically we shall propose a novel two Higgs doublet model to demonstrate this original idea. In our model, after a real Higgs singlet spontaneously breaks a special mirror symmetry, the two Higgs doublets can acquire their nearly degenerate vacuum expectation values (VEVs). At the same time, the neutrino masses and the charged fermion masses can be proportional to the difference and sum of the VEVs of two Higgs doublets, respectively. Finally we shall prove that from a phenomenological perspective, such nearly degenerate Higgs doublets with large cancellation are equivalent to a Dirac seesaw mechanism with high suppression.

\section{Mirror symmetry}

The Higgs scalars include two doublets and one real singlet, i.e.
\begin{eqnarray}
\phi_{1,2}^{}\left(\!\begin{array}{l}1,2,+\frac{1}{2}\end{array}\!\right) = \left[\begin{array}{l}\phi^{+}_{1,2}\\
[3mm]
\phi^{0}_{1,2}\end{array}\right],~~\sigma\!\left(\!\begin{array}{l}1,1,0\end{array}\!\right).
  \end{eqnarray}
Here and thereafter the brackets following the fields describe the transformations under the SM $SU(3)_c^{} \times SU(2)^{}_{L}\times U(1)_Y^{}$ gauge groups. The two Higgs doublets further respect a mirror symmetry $M_{12}^{}$, under which the real Higgs singlet performs an odd transformation, i.e. 
\begin{eqnarray}
\label{mirrorscalar}
 \phi_{1}^{}\stackrel{M_{12}^{}}{\leftarrow\!\!\!-\!\!\!-\!\!\!-\!\!\!\rightarrow} \phi_{2}^{} \,, \quad \sigma \stackrel{M_{12}^{}}{\leftarrow\!\!\!-\!\!\!-\!\!\!-\!\!\!\rightarrow} -\sigma\,.
  \end{eqnarray}
The full scalar potential then should be 
\begin{eqnarray}
\label{potential}
V&=& \frac{1}{2}\mu_\sigma^2 \sigma^2_{} + \frac{1}{4}\lambda_{\sigma}^{}\sigma^4_{} + \mu_\phi^2 \left(\phi^\dagger_1 \phi^{}_1 + \phi^\dagger_2 \phi^{}_2 \right) \nonumber\\
[2mm]
&&+\mu_{12}^2 \left(\phi^\dagger_1\phi^{}_2 +\phi^\dagger_2 \phi^{}_1\right)+ \lambda_\phi^{} \left[\left(\phi^\dagger_1 \phi^{}_1\right)^2_{} +\left( \phi^\dagger_2 \phi^{}_2 \right)^2_{}\right] \nonumber\\
[2mm]
&&+ \lambda_{12}^{}  \phi^\dagger_1 \phi^{}_1 \phi^\dagger_2 \phi^{}_2 
+  \lambda^{'}_{12}\phi^\dagger_1 \phi^{}_2 \phi^\dagger_2 \phi^{}_1 \nonumber\\
[2mm]
&&+\lambda^{''}_{12}  \left(\phi^\dagger_1 \phi^{}_1 + \phi^\dagger_2 \phi^{}_2 \right) \left(\phi^\dagger_1 \phi^{}_2 + \phi^\dagger_2 \phi^{}_1 \right) \nonumber\\
[2mm]
&&+ \lambda^{'''}_{12} \left[\left(\phi^\dagger_1\phi^{}_2\right)^2_{}+\left(\phi^\dagger_2\phi^{}_1\right)^2_{}\right] \nonumber\\
[2mm]
&&+\frac{1}{2}\lambda_{\sigma\phi}^{} \sigma^2_{}  \left(\phi^\dagger_1 \phi^{}_1 + \phi^\dagger_2 \phi^{}_2 \right)\nonumber\\
[2mm]
&&+\frac{1}{2}\lambda^{}_{\sigma  12} \sigma^2_{}  \left(\phi^\dagger_1 \phi^{}_2 + \phi^\dagger_2 \phi^{}_1\right)\nonumber\\
[2mm]
&&+ \rho_{\sigma\phi}^{}\sigma  \left(\phi^\dagger_1 \phi^{}_1 - \phi^\dagger_2 \phi^{}_2 \right) \nonumber\\
[2mm]
&&+i\rho_{\sigma 12}^{}\sigma\left(\phi^\dagger_1\phi^{}_2 - \phi^\dagger_2 \phi^{}_1\right)\,.\end{eqnarray}

In the fermion sector, we introduce three right-handed neutrinos besides three generations of the SM fermions, i.e.
\begin{eqnarray}
\label{fermion}
q_{L}^{}\!\left(\!\begin{array}{l}3,2,+\frac{1}{6}\end{array}\!\right) \!&=&\! \left[\begin{array}{l}u^{}_{L}\\
[3mm]
d^{}_{L}\end{array}\right],~~d_{R}^{}\!\left(\!\begin{array}{l}3,1,-\frac{1}{3}\end{array}\!\right),~~ u_{R}^{}\!\left(\!\begin{array}{l}3,1,+\frac{2}{3}\end{array}\!\right), \nonumber\\
[4mm]
l_{L}^{}\!\left(\!\begin{array}{l}1,2,-\frac{1}{2}\end{array}\!\right) \!&=& \!\left[\begin{array}{l}\nu^{}_{L}\\
[3mm]
e^{}_{L}\end{array}\right],~~e_{R}^{}\!\left(\!\begin{array}{l}1,1,-1\end{array}\!\right) ,~~ \nu_{R}^{}\!\left(\!\begin{array}{l}1,1,0\end{array}\!\right) .
\end{eqnarray}
Here and thereafter the family indices of fermions are omitted for simplicity. According to the mirror symmetry (\ref{mirrorscalar}), the SM fermions and the right-handed neutrinos respectively perform the even and odd transformations as below, 
\begin{eqnarray}
\label{mirrorfermion}
f_{\textrm{SM}}^{} \stackrel{M_{12}^{}}{\leftarrow\!\!\!-\!\!\!-\!\!\!-\!\!\!\rightarrow} f_{\textrm{SM}}^{} \,,\quad \nu_R^{}\stackrel{M_{12}^{}}{\leftarrow\!\!\!-\!\!\!-\!\!\!-\!\!\!\rightarrow} -\nu_R^{} \,,\end{eqnarray}
where $f_{SM}^{}$ denotes the SM fermions in Eq. (\ref{fermion}). Consequently, the allowed Yukawa couplings should be nothing but
\begin{eqnarray}
\label{yukawa}
\mathcal{L}_Y^{}&=& - y_d^{} \bar{q}_L^{} \left(\phi_1^{}+\phi_2^{}\right) d_R^{}  - y_u^{} \bar{q}_L^{} \left(\tilde{\phi}_1^{}+\tilde{\phi}_2^{}\right) u_R^{} \nonumber\\
[2mm]
&&
 -y_e^{} \bar{l}_L^{} \left(\phi_1^{}+\phi_2^{}\right) e_R^{}  -y_\nu^{} \bar{l}_L^{} \left(\tilde{\phi}_1^{} - \tilde{\phi}_2^{}\right) \nu_R^{} \nonumber\\
 [2mm]
 && +\textrm{H.c.}~~\textrm{with}~~\tilde{\phi}_{1,2}^{}\equiv i \tau_2^{} \phi_{1,2}^\ast\,.
\end{eqnarray}

Now the right-handed neutrinos are the gauge singlets so that they in principle can have a Majorana mass term. We may introduce a $U(1)_{B-L}^{}$ gauge symmetry to forbid the Majorana masses of right-handed neutrinos, meanwhile, strengthen the motivation of right-handed neutrinos as long as the related Higgs scalar has a proper $U(1)_{B-L}^{}$ charge. For example, when these SM-singlet right-handed neutrinos $v_R^{}$ have the $U(1)_{B-L}^{}$ charge $-1$ as usual, we can consider a SM-singlet Higgs scalar $\xi$ carrying an arbitrary $U(1)_{B-L}^{}$ charge except for $ \pm 2$ to spontaneously break the $U(1)_{B-L}^{}$ gauge symmetry. In consequence, the right-handed neutrinos $\nu_R^{}$ are not allowed to have the Yukawa couplings with the Higgs scalar $\xi$ so that they can not obtain any Majorana masses through the $U(1)_{B-L}^{}$ gauge symmetry breaking.

\section{Physical scalars}

The two Higgs doublets $\phi_{1,2}^{}$ and the real Higgs singlet $\sigma$ are expected to develop their VEVs, i.e. 
\begin{eqnarray}
\label{vev}
\phi^{}_{1,2} = \left[\begin{array}{c} \phi^{+}_{1,2}\\
[3mm]
\frac{1}{\sqrt{2}}\left(v_{1,2}^{}+S_{1,2}^{}+i P_{1,2}^{}\right)\end{array}\right]\,,\quad 
\sigma = v_\sigma^{}+ S_\sigma^{}\,.~~
\end{eqnarray}
Here and thereafter simply require the CP to be conserved in the scalar potential (\ref{potential}), i.e. the parameter $\rho_{\sigma12}^{}$ has been assumed a zero value.

Clearly, three would-be-Goldstone bosons, 
 \begin{eqnarray}
G_W^{\pm}&=&\frac{v_{1}^{}\phi^{\pm}_{1} +v_{2}^{} \phi^{\pm}_{2} }{\sqrt{v_{1}^2 + v_{2}^2}}\,,\\
[2mm]
G_Z^{}&=&\frac{v_{1}^{}P^{}_{1} +v_{2}^{} P^{}_{2} }{\sqrt{v_{1}^2 + v_{2}^2}}\,,
\end{eqnarray}
get eaten by the longitudinal components of the SM gauge bosons $W^{\pm}_{}$ and $Z$. Therefore, besides a pair of massive charged scalars,
\begin{eqnarray}
\label{charged}
H^{\pm}_{}&=&\frac{v_{1}^{}\phi^{\pm}_{2} - v_{2}^{} \phi^{\pm}_{1} }{\sqrt{v_{\phi_1}^2 + v_{\phi_2}^2}}\,,
\end{eqnarray}
we eventually obtain four massive neutral scalars including one pseudo scalar and three scalars, i.e. 
\begin{eqnarray}
\label{pseudo}
&&P=\frac{v_1^{}P^{}_{2} -v_2^{}P^{}_{1} }{\sqrt{v_1^2+v_2^2}}\,;\\
[2mm]
\label{scalar}
&&S_{\sigma}^{}\,,~~S_{1,2}^{}\,.
\end{eqnarray}

We then demonstrate the mass spectrum of physical scalars. For this purpose, we insert the VEVs in Eq. (\ref{vev})  into the scalar potential (\ref{potential}) and then read 
\begin{eqnarray}
\label{potential2}
V&=& \frac{1}{2}\mu_\sigma^2v_ \sigma^2+ \frac{1}{4}\lambda_{\sigma}^{}v_\sigma^4 + \frac{1}{2}\mu_\phi^2 \left(v^{2}_1 + v^{2}_2 \right) +\mu_{12}^2 v^{}_1 v^{}_2\nonumber\\
[2mm]
&&+ \frac{1}{4}\lambda_\phi^{} \left(v^4_1 + v^4_2  \right) +\left( \frac{1}{4}\lambda_{12}^{} 
+  \frac{1}{4} \lambda^{'}_{12}+  \frac{1}{2}  \lambda^{'''}_{12} \right)v_1^2 v_2^2 \nonumber\\
[2mm]
&&+\frac{1}{2}\lambda^{''}_{12}  \left(v_1^2 +v_2^2 \right)v_1^{} v_2^{}+\frac{1}{4}\lambda_{\sigma\phi}^{} v_\sigma^2  \left(v_1^2 + v_2^2 \right)\nonumber\\
[2mm]
&&+\frac{1}{2}\lambda^{}_{\sigma 12} v_\sigma^2   v_1^{}  v_2^{} +\frac{1}{2} \rho_{\sigma\phi}^{}v_\sigma^{}  \left(v_1^2- v_2^2 \right) \,.\end{eqnarray}
By minimizing the above potential, we obtain the extremal conditions as follows,
\begin{eqnarray}
\frac{\partial V}{\partial v_\sigma^{}}&=&\mu_\sigma^2 v_\sigma^{} + \lambda_\sigma^{} v_\sigma^3 + \frac{1}{2} \lambda_{\sigma\phi}^{} v_\sigma^{} \left(v_1^2 +v_2^2\right)  \nonumber\\
[2mm]
&&+ \lambda^{}_{\sigma 12} v_\sigma^{}  v_1^{}  v_2^{} +\frac{1}{2}\rho_{\sigma\phi}^{}\left(v_1^2 -v_2^2\right)\nonumber\\
[2mm]
&=&0\,,\\
[2mm]\frac{\partial V}{\partial v_1^{}}&=&\left(\mu_\phi^2 +\frac{1}{2} \lambda_{\sigma\phi}^{}v_\sigma^2\right) v_1^{} + \lambda_\phi^{} v_1^3 \nonumber\\
[2mm]
&&+\left(\frac{1}{2}\lambda_{12}^{}+\frac{1}{2}\lambda_{12}^{'}+\lambda_{12}^{'''}\right)v_1^{} v_2^2 +\frac{3}{2}\lambda_{12}^{''} v_1^2 v_2^{} \nonumber\\
[2mm]
&&+ \frac{1}{2}\lambda_{12}^{''} v_2^3 + \mu_{12}^2 v_2^{} + \frac{1}{2}\lambda^{}_{\sigma 12}v_\sigma^2 v_2^{}+ \rho_{\sigma\phi}^{} v_\sigma^{} v_1^{} \nonumber\\
[2mm]
&=& 0 \,,\\
[2mm]
\frac{\partial V}{\partial v_2^{}}&=&\left(\mu_\phi^2 +\frac{1}{2} \lambda_{\sigma\phi}^{}v_\sigma^2\right) v_2^{} + \lambda_\phi^{} v_2^3 \nonumber\\
[2mm]
&&+\left(\frac{1}{2}\lambda_{12}^{}+\frac{1}{2}\lambda_{12}^{'}+\lambda_{12}^{'''}\right)v_2^{} v_1^2 +\frac{3}{2}\lambda_{12}^{''} v_2^2 v_1^{}\nonumber\\
[2mm]
&& + \frac{1}{2}\lambda_{12}^{''} v_1^3 + \mu_{12}^2 v_1^{} + \frac{1}{2}\lambda^{}_{\sigma 12}v_\sigma^2 v_1^{}- \rho_{\sigma\phi}^{} v_\sigma^{} v_2^{}  \nonumber\\
[2mm]
&=&0 \,.\end{eqnarray}

With the above extremal conditions, the charged scalar (\ref{charged}) and the pseudo scalar (\ref{pseudo}) acquire their squared masses, respectively,
\begin{eqnarray}
m_{H^\pm}^2& =& -\left[\frac{\mu_{12}^2 +\frac{1}{2}\lambda^{}_{\sigma 12} v_\sigma^2+\frac{1}{2}\lambda_{12}^{''}\left(v_1^2+v_2^2\right)}{v_1^{}v_2^{}}\right.\nonumber\\
[2mm]
&&\left.+\frac{1}{2}\lambda^{'}_{12}+\lambda^{'''}_{12}\right] \left(v_1^2+v_2^2\right)\,,\\
[2mm]
m_{P}^2 &=& -\left[\frac{\mu_{12}^2 +\frac{1}{2}\lambda^{}_{\sigma 12} v_\sigma^2 + \frac{1}{2}\lambda^{''}_{12}\left(v_1^2+v_2^2\right)}{v_1^{}v_2^{}}\right.\nonumber\\
[2mm]&&\left.+2\lambda^{'''}_{12}\right] \left(v_1^2+v_2^2\right)\,.
\end{eqnarray}
As for the three scalars (\ref{scalar}), they have the following mass-squared matrix, 
\begin{eqnarray}
\!\!\!\!&&\left[\begin{array}{lll} m_{S_\sigma}^2 & m_{\sigma 1}^2 & m_{\sigma 2}^2\\
[3mm]
m_{\sigma 1}^2 & m_{S_1}^2 & m_{12}^2 \\ 
[3mm]
m_{\sigma 2}^2 & m_{12}^2 & m_{S_2}^2 \end{array}\right]\quad \textrm{with}\nonumber\\
[3mm]
\!\!\!\!&&m_{S_\sigma}^2 = 2\lambda_\sigma^{} v_\sigma^2 - \frac{1}{2} \rho_{\sigma\phi}^{} \left(v_1^2 - v_2^2\right)/v_\sigma^{}\,,\\
[2mm]
\!\!\!\!&&m_{\sigma 1}^2 =\lambda_{\sigma\phi}^{} v_\sigma^{} v_1^{} +\frac{1}{2}\lambda^{}_{\sigma 12} v_\sigma^{} v_1^{} + \rho_{\sigma\phi}^{} v_1^{}\,,\\
[2mm]
\!\!\!\!&&m_{\sigma 2}^2 =\lambda_{\sigma\phi}^{} v_\sigma^{} v_2^{}  +\frac{1}{2}\lambda^{}_{\sigma 12} v_\sigma^{} v_2^{}  - \rho_{\sigma\phi}^{} v_2^{}\,,\\
[2mm]
\!\!\!\!&& m_{S_1}^2 = -\left(\mu_{12}^2+\frac{1}{2}\lambda^{}_{\sigma 12} v_\sigma^2\right) \frac{v_2^{}}{v_1^{}}+ 2\lambda_\phi^{} v_1^2 +\frac{3}{2}\lambda^{''}_{12} v_1^{}v_2^{} \,,~~\nonumber\\
&&\\
[2mm]
\!\!\!\!&& m_{S_2}^2 =  -\left(\mu_{12}^2+\frac{1}{2}\lambda^{}_{\sigma 12} v_\sigma^2\right) \frac{v_1^{}}{v_2^{}}+ 2\lambda_\phi^{} v_2^2 +\frac{3}{2}\lambda^{''}_{12} v_1^{}v_2^{} \,,~~\nonumber\\
&&\\
[2mm]
\!\!\!\!&&
m_{12}^{2}=\mu_{12}^2+\frac{1}{2}\lambda^{}_{\sigma 12} v_\sigma^2 +\left(\lambda_{12}^{}+\lambda^{'}_{12}+2\lambda^{'''}_{12} \right)v_1^{} v_2^{} \nonumber\\
[2mm]
\!\!\!\!&&\quad\quad\quad+\frac{3}{2}\lambda^{''}_{12}\left(v_1^2+v_2^2\right)\,.
\end{eqnarray}
Here we have taken the base $(S_\sigma^{}, ~S_1^{}, ~S_2^{})^T_{}$. The above matrix can be diagonalized in principle. For simplicity we do not perform this diagonalization in the present work.

\section{Fermion masses}

Through their Yukawa couplings with the Higgs doublets, i.e. Eq. (\ref{yukawa}), the charged fermions and the neutral neutrinos obtain the Dirac masses as below,  
\begin{eqnarray}
\mathcal{L}&\supset& - m_d^{} \bar{d}_L^{}d_R^{}  - m_u^{} \bar{u}_L^{} u_R^{}-m_e^{} \bar{e}_L^{} e_R^{}  -m_\nu^{} \bar{\nu}_L^{} \nu_R^{} \nonumber\\
[2mm]
&&+\textrm{H.c.}\quad \textrm{with} \nonumber\\
[2mm]
&&m_d^{}=\frac{1}{\sqrt{2}}y_d^{}\left(v_1^{}+v_2^{}\right) \,,\\
[2mm]
&&m_u^{}=\frac{1}{\sqrt{2}}y_u^{}\left(v_1^{}+v_2^{}\right) \,,\\
[2mm]
&&m_e^{}=\frac{1}{\sqrt{2}}y_e^{}\left(v_1^{}+v_2^{}\right)\,,\\
[2mm]
&&m_\nu^{}=\frac{1}{\sqrt{2}}y_\nu^{}\left(v_1^{}-v_2^{}\right)\,.
\end{eqnarray}
Remarkably, the charged fermion masses and the neutrino masses are respectively proportional to the sum and difference of the VEVs of two Higgs doublets. This means the neutrino masses would become zero if the two Higgs doublets had same VEVs. Fortunately, the degeneracy between two Higgs doublets will be spontaneously violated when the real Higgs singlet develops its VEV \cite{bm1989} to spontaneously break the mirror symmetry defined in Eqs. (\ref{mirrorscalar}) and (\ref{mirrorfermion}). Alternatively, we may softly break this mirror symmetry in the scalar potential in order to simply avoid the possible problem of domain wall.

Needless to say, we are interested in naturally generating the tiny neutrino masses. So, the difference of the VEVs of two Higgs doublets is expected to be small enough so that the related Yukawa couplings can arrive at a sizeable level. In this sense, we would like to understand that the smallness of neutrino masses is induced by a large cancellation between the VEVs of two Higgs doublets.

For demonstration, we denote
\begin{eqnarray}
v_1^{}-v_2^{} \ll v_{1}^{}\simeq v_{2}^{}\simeq v_{12}^{}~~\textrm{with}~~v_{12}^{}\equiv \frac{1}{2}\left(v_1^{}+v_2^{}\right)\,.~~
\end{eqnarray}
It is easy to find that the VEVs of two Higgs doublets can have a good approximation as below,  
\begin{eqnarray}
\label{vev1}
v_1^{2}\simeq v_2^{2}& \simeq&-\frac{\mu_{12}^2 +\frac{1}{2} \lambda_{\sigma 12}^{} v_\sigma^2+\mu_{\phi}^2 +\frac{1}{2} \lambda_{\sigma\phi}^{} v_\sigma^2}{\lambda_{\phi}^{} +\frac{1}{2}\lambda_{12}^{} + \frac{1}{2}\lambda_{12}^{'}+ 2\lambda_{12}^{''}+ \lambda_{12}^{'''}}\,,\\
[2mm]
\label{vev2}
v_1^{} - v_2^{}&\simeq & 2\rho_{\sigma\phi}^{} v_\sigma^{} v_{12}^{}/\left[2\mu_{12}^2 + \lambda^{}_{\sigma 12} v_\sigma^2\right.\nonumber\\
[2mm]
&&\left.-\left(2\lambda_\phi^{}-\lambda_{12}^{} -\lambda_{12}^{'} -2\lambda_{12}^{''}- 2\lambda_{12}^{'''}\right)v_{12}^2 \right]\,.\nonumber\\
&&
\end{eqnarray}
Obviously, the difference $v_1^{} - v_2^{}$ definitely would disappear for a zero value of the dimensional coupling $\rho_{\sigma\phi}^{}$ or the VEV $v_\sigma^{}$.

In order to guarantee the difference $v_1^{} - v_2^{}$ to be small but nonzero, we prefer to take $\rho_{\sigma\phi}^{}$ small enough while keep $v_\sigma^{}$ large enough. For such parameter choice, the scalar $S_\sigma^{}$ from the real Higgs singlet $\sigma$ can be heavy enough to avoid possible BBN constraint and other experimental limits, meanwhile, the scalars $H^{\pm}_{}$, $S_{1,2}^{}$ and $P$ from the two Higgs doublets $\phi_{1,2}^{}$ can be light enough to predict rich collider phenomena and other experimental signals \cite{dl2009}. We shall study the phenomenological details elsewhere.

\section{Dirac seesaw}

Inspired by the structure of the Yukawa couplings (\ref{yukawa}), we try to illustrate our scenario from another framework based on the following linear combinations of two Higgs doublets, 
\begin{eqnarray}
\varphi_1^{}=\frac{1}{\sqrt{2}}\left(\phi^{}_1 + \phi^{}_2\right)\,,\quad \varphi_2^{}=\frac{1}{\sqrt{2}}\left(\phi^{}_{1}-\phi^{}_{2}\right)\,.
\end{eqnarray}
The Yukawa couplings (\ref{yukawa}) then become to be
\begin{eqnarray}
\label{yukawa2}
\mathcal{L}_Y^{}&=& -\sqrt{2}y_d^{} \bar{q}_L^{} \varphi_1^{} d_R^{}  -\sqrt{2}y_u^{} \bar{q}_L^{} \tilde{\varphi}_1^{}u_R^{} 
 -\sqrt{2}y_e^{} \bar{l}_L^{} \varphi_1^{} e_R^{} \nonumber\\
 [2mm]
 && -\sqrt{2}y_\nu^{} \bar{l}_L^{} \tilde{\varphi}_2^{}\nu_R^{} +\textrm{H.c.}~~\textrm{with}~~\tilde{\varphi}_{1,2}^{}\equiv i \tau_2^{} \varphi_{1,2}^\ast\,.~~~~
\end{eqnarray}
Meanwhile, the scalar potential (\ref{potential}) is rewritten by 
\begin{eqnarray}
\label{potential3}
V&=& \frac{1}{2}\mu_\sigma^2 \sigma^2_{} + \frac{1}{4}\lambda_{\sigma}^{}\sigma^4_{} + \mu_1^2 \varphi^\dagger_{1} \varphi^{}_{1} + \mu_{2}^2 \varphi^\dagger_{2} \varphi^{}_{2}\nonumber\\
[2mm]
&&+\lambda_1^{} \left(\varphi^\dagger_{1}\varphi_1^{}\right)^2_{}  +\lambda_2^{}\left(\varphi^\dagger_{2}\varphi^{}_2 \right)^2_{} +\lambda^{}_{3} \varphi^\dagger_{1}\varphi^{}_1 \varphi^\dagger_{2}\varphi^{}_2 \nonumber\\
[2mm]
&&+\lambda^{}_{4} \varphi^\dagger_{1} \varphi^{}_2 \varphi^\dagger_{2} \varphi^{}_1 +\lambda^{}_{5}\left[\left( \varphi^\dagger_{1}\varphi^{}_2\right)^2_{}+\left(\varphi^\dagger_{2}\varphi^{}_1\right)^2_{}\right]  \nonumber\\
[2mm]
&&+\frac{1}{2} \lambda_{\sigma\phi}^{}\sigma^2_{}\left(\varphi^\dagger_{1}\varphi^{}_1+\varphi^\dagger_{2}\varphi^{}_2\right)  \nonumber\\
[2mm]
&&+\frac{1}{2} \lambda_{\sigma12}^{}\sigma^2_{}\left(\varphi^\dagger_{1}\varphi^{}_1-\varphi^\dagger_{2}\varphi^{}_2\right)  \nonumber\\
[2mm]
&&+ \rho_{\sigma\varphi}^{}\sigma  \varphi^\dagger_1\varphi^{}_2 + \rho_{\sigma\varphi}^\ast \sigma \varphi^\dagger_2 \varphi^{}_1\,. \end{eqnarray}
Here the parameters are determined by 
\begin{eqnarray}
\mu_1^2& =& \mu_\phi^2 + \mu_{12}^2\,,\\
[2mm]
\mu_2^2 &=& \mu_\phi^2 -\mu_{12}^2\,,\\
[2mm]
\lambda_1^{}&=& \frac{1}{2}\lambda_\phi^{} + \frac{1}{4} \lambda_{12}^{} + \frac{1}{4} \lambda^{'}_{12} + \lambda^{''}_{12} + \frac{1}{2}\lambda^{'''}_{12}\,,\\
[2mm]
\lambda_2^{}&=& \frac{1}{2}\lambda_\phi^{} + \frac{1}{4} \lambda_{12}^{} + \frac{1}{4} \lambda^{'}_{12} - \lambda^{''}_{12} + \frac{1}{2}\lambda^{'''}_{12}\,,\\
[2mm]
\lambda_{3}^{}&=& \lambda_\phi^{} + \frac{1}{2} \lambda_{12}^{} - \frac{1}{2} \lambda^{'}_{12} -\lambda^{'''}_{12}\,,\\
[2mm]
\lambda_{4}^{}&=& \lambda_\phi^{} - \frac{1}{2} \lambda_{12}^{} + \frac{1}{2} \lambda^{'}_{12} -\lambda^{'''}_{12}\,,\\
[2mm]
\lambda_{5}^{}&=& \frac{1}{2}\lambda_\phi^{} - \frac{1}{4} \lambda_{12}^{} - \frac{1}{4} \lambda^{'}_{12} +  \frac{1}{2}\lambda^{'''}_{12}\,,\\
[2mm]
\rho_{\sigma\varphi}^{}&=& \rho_{\sigma\phi}^{} - i \rho_{\sigma 12}^{}\,.
 \end{eqnarray}

Now the Higgs doublet $\varphi_1^{}$ only participates in the Yukawa couplings for the charged fermion mass generation while the Higgs doublet $\varphi_{2}^{}$ only participates in the Yukawa couplings for the neutrino mass generation. Furthermore, the trilinear coupling among the two Higgs doublets $\varphi_{1,2}^{}$ and the real Higgs singlet $\sigma$ can realize a mass mixing between the two Higgs doublets $\varphi_{1,2}^{}$ after the real Higgs singlet $\sigma$ develops its VEV. Such context indeed accommodates a Dirac seesaw scenario \cite{gh2006}, where the Higgs doublet $\varphi_2^{}$ can provide rich phenomena if its charged and neutral components are not far above the TeV scale \cite{dl2009}.

Note if we initially start with the base $\varphi_{1,2}^{}$ and then write down the Yukawa couplings (\ref{yukawa2}) and the scalar potential (\ref{potential3}), we should impose a discrete symmetry to distinguish the roles of $\varphi_{1,2}^{}$. In fact, this is just a $Z_2^{}$ discrete symmetry, under which the real Higgs singlet $\sigma$, the Higgs doublet $\varphi_2^{}$ and the right-handed neutrinos $\nu_R^{}$ carry an odd parity while the Higgs doublet $\varphi_1^{}$ and the SM fermions $f_{SM}^{}$ carry an even parity, i.e.
\begin{eqnarray}
\label{sz2}
\!\!\!\!\!\!\!\!&& \varphi_{1}^{}\stackrel{Z_2^{}}{\leftarrow\!\!\!-\!\!\!-\!\!\!-\!\!\!\rightarrow} \varphi_{1}^{} \,, \quad  \varphi_{2}^{}\stackrel{Z_2^{}}{\leftarrow\!\!\!-\!\!\!-\!\!\!-\!\!\!\rightarrow}- \varphi_{2}^{} \,, \quad \sigma \stackrel{Z_2^{}}{\leftarrow\!\!\!-\!\!\!-\!\!\!-\!\!\!\rightarrow} -\sigma\,;~~~~\\
 [2mm]
 \label{fz2}
 \!\!\!\!\!\!\!\!&&f_{\textrm{SM}}^{} \stackrel{Z_{2}^{}}{\leftarrow\!\!\!-\!\!\!-\!\!\!-\!\!\!\rightarrow} f_{\textrm{SM}}^{} \,,\quad \nu_R^{}\stackrel{Z_{2}^{}}{\leftarrow\!\!\!-\!\!\!-\!\!\!-\!\!\!\rightarrow} -\nu_R^{} \,.~~~~
 \end{eqnarray}

We would like to emphasize that although the original base $\phi_{1,2}^{}$ and the new base $\varphi_{1,2}^{}$ result in the same phenomena, they should provide two different interpretations on the neutrino mass generation, i.e. one is a large cancellation while the other is a seesaw suppression. This is an inevitable reflection on two different theoretical starting points, i.e. one is the mirror symmetry $M_{12}^{}$ defined by Eqs. (\ref{mirrorscalar}) and (\ref{mirrorfermion}), while the other is the discrete symmetry $Z_2^{}$ defined by Eqs. (\ref{sz2}) and (\ref{fz2}). Note that the mirror symmetry $M_{12}^{}$ can not be trivially identified with the discrete symmetry $Z_2^{}$. Actually,  unlike the mirror symmetry $M_{12}^{}$, the discrete symmetry $Z_2^{}$ can not constrain the number of Higgs doublets if it is initially introduced, i.e. the number of $Z_2^{}$-even Higgs doublets can be totally different from the number of $Z_2^{}$-odd Higgs doublets.

\section{Conclusion}

In the present work we have explored a novel scenario where the smallness of neutrino masses can be naturally understood by a large cancelation. Specifically we construct a mirror symmetry between two Higgs doublets. Then the right-handed neutrinos perform an odd transformation under this mirror symmetry while the SM fermions perform an even transformation. As a result, the difference and sum of two Higgs doublets respectively take part in the Yukawa couplings for generating the neutrino masses and the charged fermion masses. After the mirror symmetry is spontaneously broken by a real Higgs singlet, the VEVs of two Higgs doublets can acquire a small difference. Accordingly, the neutrino masses and the charged fermion masses respectively are proportional to the difference and sum of the VEVs of two Higgs doublets. We also clarify that the present large cancellation can be phenomenologically equivalent to a Dirac seesaw mechanism although their theoretical starting points are absolutely different.

\textbf{Acknowledgement}: This work was supported in part by the National Natural Science Foundation of China under Grant No. 12175038 and in part by the Fundamental Research Funds for the Central Universities.

\end{document}